\documentclass[twoside,11pt]{article}
\usepackage{amsmath}
\usepackage{stackrel}
\usepackage{algorithm}
\usepackage{jmlr2e}

\usepackage[colorlinks = true,
            linkcolor = blue,
            urlcolor  = blue,
            citecolor = blue,
            anchorcolor = blue]{hyperref}
            
\usepackage[noend]{algpseudocode}
\makeatletter
\def\BState{\State\hskip-\ALG@thistlm}
\makeatother

\usepackage{listings}
\usepackage{color}

\definecolor{dkgreen}{rgb}{0,0.6,0}
\definecolor{gray}{rgb}{0.5,0.5,0.5}
\definecolor{mauve}{rgb}{0.58,0,0.82}

\ShortHeadings{A reparameterization of Ridge Regression}{Rokem and Kay}
\firstpageno{1}

\begin{document}

\title{Fractional ridge regression: a fast, interpretable reparameterization of ridge regression}

\author{\name Ariel Rokem \email arokem@uw.edu \\
       \addr The University of Washington eScience Institute\\
       University of Washington\\
       Seattle, WA 98195, USA
       \AND
       \name Kendrick Kay \email kay@umn.edu \\
       \addr Center for Magnetic Resonance Research \\
       University of Minnesota\\
       Twin Cities, MN 55455, USA}

\maketitle

\begin{abstract}

Ridge regression (RR) is a regularization technique that penalizes the L2-norm of the coefficients in linear regression. One of the challenges of using RR is the need to set a hyperparameter ($\alpha$) that controls the amount of regularization. Cross-validation is typically used to select the best $\alpha$ from a set of candidates. However, efficient and appropriate selection of $\alpha$ can be challenging, particularly where large amounts of data are analyzed. Because the selected $\alpha$ depends on the scale of the data and predictors, it is not straightforwardly interpretable. Here, we propose to reparameterize RR in terms of the ratio $\gamma$ between the L2-norms of the regularized and unregularized coefficients. This approach, called fractional RR (FRR), has several benefits: the solutions obtained for different $\gamma$ are guaranteed to vary, guarding against wasted calculations, and automatically span the relevant range of regularization, avoiding the need for arduous manual exploration. We provide an algorithm to solve FRR, as well as open-source software implementations in Python and MATLAB (\url{https://github.com/nrdg/fracridge}). We show that the proposed method is fast and scalable for large-scale data problems, and delivers results that are straightforward to interpret and compare across models and datasets.

\end{abstract}


\newpage 

\section{Introduction}
Consider the standard linear model setting $Y=X\beta$ solved for $\beta$, where $Y$ is the $(d,~t)$ data matrix ($d$ data points in each of $t$ targets), $X$ is the $(d,~p)$ design matrix ($d$ data points for each of $p$ predictors), and $\beta$ is a $(p,~t)$ matrix (with $p$ coefficients, one for each predictor, for each of the targets). Ordinary least-squares regression (OLS) and regression based on the Moore-Penrose pseudoinverse (in cases where $p > d$) attempt to find the set of coefficients $\beta$ that minimize squared error for each of the targets $y$. While these unregularized approaches have some desirable properties, in practical applications where noise is present, they tend to overfit the coefficient parameters to the noise present in the data. Moreover, they tend to cause unstable parameter estimates in situations where predictors are highly correlated.

Ridge regression \citep{hoerl1970ridge} addresses these issues by trading off the addition of some bias for the reduction of eventual error (e.g., measured using cross-validation \citep{stone1978cross, stone1974cross}). It does so by penalizing not only the sum of the squared errors in fitting the data for each target, but by also minimizing the squared L2-norm of the solution, $||\beta||_2^2 = \sum{\beta^2}$. Fortunately, this form of regularization does not incur a substantial computational cost. This is because it can be implemented using the same numerical approach for solving unregularized regression, with the simple addition of a diagonal matrix $\alpha I$ to the standard matrix equations. Thus, the computational cost of solving ridge regression is essentially identical to that of the unregularized solution. Thanks to its simplicity, computational expedience, and its robustness in different data regimes, ridge regression is a very popular technique, with the classical references describing the method \citep{hoerl1970ridge, tikhonov1977solutions} cited more than 25,000 times according to Google Scholar. 

However, beneath the apparent simplicity of ridge regression is the fact that for most applications, it is impossible to determine \emph{a priori} the degree of regularization that yields the best solution. This means that in typical practice, researchers must test several different hyperparameter values $\alpha$ and select the one that yields the least cross-validation error on a set of data specifically held out for hyperparameter selection. In large-scale data problems, the number of data points $d$, number of predictors $p$, and/or number of targets $t$ can be quite large. This has the consequence that the number of hyperparameter values that are tested, $f$, can pose a prohibitive computational barrier.

Given the difficulty of predicting the effect of $\alpha$ on solution outcomes, it is common practice to test values that are widely distributed on a log scale (for example, see \citep{friedman2010regularization}). Although this approach is not grounded in a particular theory, as long as the values span a large enough range and are spaced densely enough, an approximate minimum of the cross-validation error is likely to be found. But testing many $\alpha$ values can be quite costly, and the practitioner might feel tempted to cull the set of values tested. In addition, it is always a possibility that the initial chosen range might be mismatched to the problem at hand. Sampling $\alpha$ values that are too high or too low will produce non-informative candidate solutions that are either over-regularized ($\alpha$ too high) or too similar to the unregularized solution ($\alpha$ too low). Thus, in practice, conventional implementations of ridge regression may produce poor solutions and/or waste substantial computational time. 

Here, we propose a simple reparameterization of ridge regression that overcomes the aforementioned challenges. Our approach is to produce coefficient solutions that have an L2-norm that is a pre-specified fraction of the L2-norm of the unregularized solution. In this approach, called \emph{fractional ridge regression} (FRR), redundancies in candidate solutions are avoided because solutions with different fractional L2-norms are guaranteed to be different. Moreover, by targeting fractional L2-norms that span the full range from 0 to 1, the FRR approach explores the full range of effects of regularization on $\beta$ values from under- to over-regularization, thus assuring that the best possible solution is within the range of solutions explored. We provide a fast and automated algorithm to calculate FRR, and provide open-source software implementations in Python and MATLAB. We demonstrate in benchmarking simulations that FRR is computationally efficient for even extremely large data problems, and we show that FRR applies successfully to real-world data and delivers clear and interpretable results. Overall, FRR may prove particularly useful for researchers tackling large-scale datasets where automation, efficiency, and interpretability are critical.

\section{Methods}

\subsection{Background and theory}
\label{sec:background}

Consider the dataset $y$ with dimensionality $d$ (number of data points) by $t$ (number of targets). Each column in $y$ represents a separate target for linear regression: 

\begin{equation} \label{eq:regression}
y = X\beta + \epsilon
\end{equation}

where $y$ is the measured data for a single target (dimensionality $d$ by $1$; we drop the index on y for notational convenience), $X$ is the ``design'' matrix with predictors (dimensionality $d$ by $p$), $\beta$ are the coefficients (dimensionality $p$ by 1), and $\epsilon$ is a noise term. Our typical objective is to solve for $\beta$ in a way that minimizes the squared error. If $X$ is full rank, the ordinary least squares (OLS) solution to this problem is: 

\begin{equation} \label{eq:OLS}
\hat{\beta}^{OLS} = (X^t X)^{-1} X^t y
\end{equation}

In cases where $X$ is not full rank, the OLS solution is no longer well-defined and the Moore-Penrose pseudoinverse is used instead. We will refer to these unregularized approaches collectively as OLS.

To regularize the OLS solution, ridge regression applies a penalty to the squared L2-norm of the coefficients, leading to a different estimator for $\beta$: 

\begin{equation} \label{eq:rr}
\hat{\beta}^{RR} = (X^t X + \alpha I)^{-1} X^t y
\end{equation}

where $\alpha$ is a hyperparameter and $I$ is the identity matrix \citep{hoerl1970ridge, tikhonov1977solutions}. For computational efficiency, it is well known that the original problem can be rewritten \citep{Hastie2004-nd} using singular value decomposition (SVD) of the matrix $X$:

\begin{equation} \label{eq:svd}
X = U S V^t
\end{equation}

with $dim(U) = (d, p)$, $dim(S) = (p, p)$ and $dim(V) =  (p, p)$ .

Note that $S$ is a square matrix: 

\[
   S=
  \left[ {\begin{array}{ccccc}
   \lambda_1 & 0 & ... &  \\
   0 & \lambda_2 & 0 & ...  \\
   0 & 0 & \lambda_3 & 0 & ... \\
   \vdots \\
   ... & 0 & 0 & 0 & \lambda_p \\
  \end{array} } \right]
\]

with $\lambda_i$ as the singular values ordered from largest to smallest. Replacing the design matrix $X$ with its SVD, we obtain:

\begin{equation} \label{eq:svd2}
y = U S V^t \beta + \epsilon.
\end{equation}

Given that $U$ is unitary (i.e., $U^tU$ is $I$), left-multiplying each side with $U^t$ produces:

\begin{equation} \label{eq:svd_rotation}
U^ty = SV^t \beta + U^t \epsilon.
\end{equation}

Let $\tilde{y} = U^ty$, $\tilde{\beta} = V^t \beta$, and $\tilde{\epsilon} = U^t\epsilon$. These are transformations (rotations) of the original quantities ($y$, $\beta$, and $\epsilon$) through the unitary matrices $U^t$ and $V^t$. In cases where $p < d$, this also projects the quantities into a lower-dimensional space of dimensionality $p$. The OLS solution can be obtained in this space:

\begin{equation} \label{eq:ols_svd}
\tilde{\hat{\beta}}^{OLS} = (S^t S)^{-1} S^t \tilde{y}, 
\end{equation}
which simplifies to:

\begin{equation} \label{eq:ols_svd2}
\tilde{\hat{\beta}}^{OLS} = S^{-2} (\lambda \odot \tilde{y}), 
\end{equation}where $\odot$ is the Hadamard (element-wise) product, and 

\[
   S^{-2}=
  \left[ {\begin{array}{ccccc}
   \frac{1}{\lambda_1^2} & 0 & ... &  \\
   0 & \frac{1}{\lambda_2^2} & 0 & ...  \\
   0 & 0 & \frac{1}{\lambda_3^2} & 0 & ... \\
   \vdots \\
   ... & 0 & 0 & 0 &  \frac{1}{\lambda_p^2}\\
  \end{array} } \right]
\]is the inverse of the square of the singular value matrix $S$. Thus, in the lower-dimensional space, we can solve the OLS problem with a scalar multiplication: 

\begin{equation} \label{eq:ols_coeff}
\tilde{\hat{\beta}}^{OLS}_i = \frac{1}{\lambda_i^2} \lambda_i \tilde{y_i}, 
\end{equation}

which simplifies finally to

\begin{equation} \label{eq:ols_solution_ut}
\tilde{\hat{\beta}}^{OLS}_i = \frac{\tilde{y_i}}{\lambda_i}.
\end{equation}

The SVD-based reformulation of regression described above is additionally useful as it provides insight into the nature of ridge regression \citep{skouras1994estimation}. Specifically, consider the ridge regression solution in the low-dimensional space: 

\begin{equation} \label{eq:rr_in_u_space}
\tilde{\hat{\beta}}^{RR} = (S^t S + \alpha I)^{-1} S^t \tilde{y}
\end{equation}

To compute this solution, we note that: 

\begin{equation} \label{eq:xxx}
S^tS + \alpha I = 
  \left[ {\begin{array}{ccccc}
   \lambda_1^2 + \alpha & 0 & ... &  \\
   0 & \lambda_2^2 + \alpha& 0 & ...  \\
   0 & 0 & \lambda_3^2 + \alpha& 0 & ... \\
   \vdots \\
   ... & 0 & 0 & 0 & \lambda_p^2 + \alpha \\
  \end{array} } \right]
\end{equation}

the inverse of which is: 

\begin{equation} \label{eq:svd_rr_matrix}
(S^tS + \alpha I)^{-1} = 
  \left[ {\begin{array}{ccccc}
   \frac{1}{\lambda_1^2 + \alpha} & 0 & ... &  \\
   0 & \frac{1}{\lambda_2^2 + \alpha} & 0 & ...  \\
   0 & 0 & \frac{1}{\lambda_3^2 + \alpha}& 0 & ... \\
   \vdots \\
   ... & 0 & 0 & 0 & \frac{1}{\lambda_p^2 + \alpha} \\
  \end{array} } \right]
\end{equation}

Finally, plugging into equation \ref{eq:rr_in_u_space}, we obtain:

\begin{equation} \label{eq:rr_coeff}
\tilde{\hat{\beta}}^{RR}_i = \frac{\lambda_i}{\lambda_i^2 + \alpha}  \tilde{y_i}
\end{equation}

This shows that in the low-dimensional space, ridge regression can be solved using scalar operations.

To further illustrate the relationship between the ridge regression and OLS solutions, by plugging equation \ref{eq:ols_solution_ut} into equation \ref{eq:rr_coeff}, we observe the following:

\begin{equation} \label{eq:shrinkage_ut}
\tilde{\hat{\beta}}^{RR}_i = \frac{\lambda_i^2}{\lambda_i^2 + \alpha} \tilde{\hat{\beta_i}}^{OLS}
\end{equation}In other words, the ridge regression coefficients are simply scaled-down versions of the OLS coefficients, with a different amount of shrinkage for each coefficient. Coefficients associated with larger singular values are less shrunken than those with smaller singular values.

To obtain solutions in the original space, it is necessary to left-multiply the coefficients with $V$:

\begin{equation} \label{eq:orig_coeff}
\hat{\beta} = V\tilde{\hat{\beta}} 
\end{equation}



Next, we turn to fractional ridge regression (FRR). The core concept of FRR is to reparameterize ridge regression in terms of the amount of shrinkage applied to the overall L2-norm of the solution. Specifically, we define the fraction $\gamma$ as: 

\begin{equation}
\label{eq:gamma}
\gamma =  \frac{||\tilde{\hat{\beta}}^{RR}||_2}{||\tilde{\hat{\beta}}^{OLS}||_2} 
\end{equation}
Because $V$ is a unitary transformation, the L2-norm of a coefficient solution in the low-dimensional space, $||\hat{\tilde{\beta}}||_2$, is identical to the L2-norm of the coefficient solution in the original space, $||\hat{\beta}||_2$. Thus, we can operate fully within the low-dimensional space and be guaranteed that the fractions will be maintained in the original space.

In FRR, instead of specifying desired values for $\alpha$, we instead specify values of $\gamma$ between 1 (no regularization) and 0 (full regularization). But how can one compute the ridge regression solution for a specific desired value of $\gamma$? Based on equations \ref{eq:ols_coeff} and \ref{eq:rr_coeff}, it is easy to calculate the value of $\gamma$ corresponding to a specific given $\alpha$ value:

\begin{equation}
\label{eq:overall_shrinkage}
\gamma =  \frac{||\tilde{\hat{\beta}}^{RR}||_2}{||\tilde{\hat{\beta}}^{OLS}||_2} = \sqrt{\frac{\sum{(\frac{\lambda_i \tilde{y}_i}{\lambda_i^2 + \alpha})^2}}{\sum{(\frac{\tilde{y}_i}{\lambda_i}})^2}}
\end{equation}

In some special cases, this calculation can be considerably simplified. For example, if the eigenvalue spectrum of $X$ is flat ($\lambda_i = \lambda_j$ for any $i \neq j$), we can set all the singular values to $\lambda$, yielding the following:

\begin{equation}
\gamma = \sqrt{\frac{ p (\frac{\lambda}{\lambda^2 + \alpha})^2 {\sum{\tilde{y}^2_i}}}{p (\frac{1}{\lambda})^2 \sum{\tilde{y_i}^2 }}}  =  
\frac{\frac{\lambda}{\lambda^2 + \alpha} {}}{\frac{1}{\lambda} }   = \frac{\lambda^2}{\lambda^2 + \alpha},
\end{equation}

This recapitulates the result obtained in \citep{hoerl1970ridge}, equation 2.6. We can then solve for $\alpha$:  

\begin{equation}
\alpha = \lambda^2 (\frac{1}{\gamma} -  1)
\end{equation}
Thus, in this case, there is an analytic solution for the appropriate $\alpha$ value, and one can proceed to compute the ridge regression solution using equation \ref{eq:rr_coeff}. Another special case is if we assume that all values of $\tilde{y}_i$ are identical. In this case, we can easily calculate the achieved shrinkage, but in terms of L1-norm (not L2-norm):

\begin{equation} 
\label{eq:l1_norm_shrinkage}
\frac{||\tilde{\hat{\beta}}^{RR}||_1}{||\tilde{\hat{\beta}}^{OLS}||_1}
= \frac{\sum{\frac{\lambda_i \tilde{y}}{\lambda_i^2 + \alpha}}}{\sum{\frac{\tilde{y}}{\lambda_i}}} =  \frac{\tilde{y}\sum{\frac{\lambda_i}{\lambda_i^2 + \alpha}}}{\tilde{y}\sum{\frac{1}{\lambda_i}}} = \sum_{i=1}^{p} \frac{\lambda_i^2}{\lambda_i^2 + \alpha}.
\end{equation}
Notice that this is the sum of the shrinkages for individual coefficients from equation \ref{eq:shrinkage_ut}, and has been defined as the \emph{effective degrees of freedom} of ridge regression (See \cite{hastie01statisticallearning}, pg. 68).

These two special cases have the appealing feature that the regularization level can be controlled on the basis of examining only the design matrix $X$. However, they rely on strong assumptions that are not guaranteed to hold in general. Thus, for accurate ridge regression outcomes, we see no choice but to develop an algorithm that uses both the design matrix $X$ and the data values $y$.

\subsection{An algorithm for solving fractional ridge regression}.
\label{sec:methods_algorithm}

Our proposed algorithm is straightforward: it evaluates $\gamma$ for a range of $\alpha$ values and uses interpolation to determine the $\alpha$ value that achieves the desired fraction $\gamma$. Although this method relies on brute force and may not seem mathematically elegant, it achieves accurate outcomes and, somewhat surprisingly, can be carried out with minimal computational cost.

The algorithm receives as input a design matrix $X$, target variables $Y$, and a set of requested fractions $\gamma$. The algorithm calculates the FRR solutions for all targets in $Y$, returning estimates of the coefficients $\hat{\beta}$ as well as the values of hyperparameter $\alpha$ that correspond to each requested $\gamma$. In the text below, we indicate the lines of code that implement each step of the algorithm (see also section \ref{sec:software} below) in the MATLAB (designated with ``M'') and Python (designated with ``P'') implementations.
\begin{enumerate}

\item Compute the SVD of the design matrix, $USV^t = X$ (\href{https://github.com/nrdg/fracridge/blob/0.4/matlab/fracridge.m#L247}{M247}, \href{https://github.com/nrdg/fracridge/blob/0.4/fracridge/fracridge.py#L73}{P73}). To avoid numerical instability, very small eigenvalues of $X$ are treated as 0.

\item The data are transformed $\tilde{y} = U^t y$  (\href{https://github.com/nrdg/fracridge/blob/0.4/matlab/fracridge.m#L254}{M254}, \href{https://github.com/nrdg/fracridge/blob/0.4/fracridge/fracridge.py#L75}{P75}). 

\item The OLS problem is solved with one broadcast division operation (equation \ref{eq:ols_solution_ut}) (\href{https://github.com/nrdg/fracridge/blob/0.4/matlab/fracridge.m#272}{M272}, \href{https://github.com/nrdg/fracridge/blob/0.4/fracridge/fracridge.py#L79}{P79}). 

\item The values of $\alpha$ that satisfy the FRR requirement are guaranteed to lie within a range that depends on the eigenvalues of $X$. A series of initial candidate values of $\alpha$ are selected to span a log-spaced range from $10^{-3} \lambda_p^2$, much smaller than the smallest singular value of the design matrix, to $10^3 \lambda_1^2$, much larger than the largest singular value of the design matrix (\href{https://github.com/nrdg/fracridge/blob/0.4/matlab/fracridge.m#L295-L299}{M295-299}, \href{https://github.com/nrdg/fracridge/blob/0.4/fracridge/fracridge.py#L89-L96}{P89-96}). Based on testing on a variety of regression problems, we settled on a spacing of $0.2~log_{10}$ units within the range of candidate $\alpha$ values. This provides fine enough gridding such that interpolation results are nearly perfect (empirical fractions are approximately $1\%$ or less from the desired fractions). 

\item  Based on equation \ref{eq:shrinkage_ut}, a scaling factor for every value of $\alpha$ and every eigenvalue $\lambda$ is calculated as (\href{https://github.com/nrdg/fracridge/blob/0.4/matlab/fracridge.m#L310-L312}{M310-312}, \href{https://github.com/nrdg/fracridge/blob/0.4/fracridge/fracridge.py#L97-L98}{P97-98}):

\begin{equation}
\label{eq:algo_scaling}
    SF_{i, j} = \lambda_i^2 / (\lambda_i^2 + \alpha_j)
\end{equation}

\item The main loop of the algorithm iterates over targets. For every target, the scaling in equation \ref{eq:algo_scaling} is applied to the computed OLS coefficients (from Step 3), and the L2-norm of the solution at each $\alpha_j$ is divided by the L2-norm of the OLS solution to determine the fractional length, $\gamma_j$ (\href{https://github.com/nrdg/fracridge/blob/0.4/matlab/fracridge.m#L332-L345}{M332-345}, \href{https://github.com/nrdg/fracridge/blob/0.4/fracridge/fracridge.py#L111-L112}{P111-112}). 

\item Interpolation is used with $\alpha_j$ and $\gamma_j$ to find values of $\alpha$ that correspond to the desired values of $\gamma$  (\href{https://github.com/nrdg/fracridge/blob/0.4/matlab/fracridge.m#L361}{M361}, \href{https://github.com/nrdg/fracridge/blob/0.4/fracridge/fracridge.py#L113}{P113}). These target $\alpha$ values are then used to calculate the ridge regression solutions via equation \ref{eq:shrinkage_ut} (\href{https://github.com/nrdg/fracridge/blob/0.4/matlab/fracridge.m#L367}{M367}, \href{https://github.com/nrdg/fracridge/blob/0.4/fracridge/fracridge.py#L116-L117}{P116-117}). 

\item After the iteration over targets is complete, the solutions are transformed to the original space by multiplying $\hat{\beta} = V \tilde{\hat{\beta}}$ (\href{https://github.com/nrdg/fracridge/blob/0.4/matlab/fracridge.m#L418}{M418}, \href{https://github.com/nrdg/fracridge/blob/0.4/fracridge/fracridge.py#L119}{PL119}).

\end{enumerate}

In summary, this algorithm requires just one (potentially computationally expensive) initial SVD of the design matrix. Operations done on a per-target basis are generally inexpensive, relying on fast vectorized array operations, with the exception of the interpolation step. Although a large range of candidate $\alpha$ values are evaluated internally by the algorithm, these values are eventually discarded, thereby avoiding costs associated with the final step (multiplication with $V$).

\subsection{Software implementation}
\label{sec:software}
We implemented the algorithm described in section \ref{sec:methods_algorithm} in two different popular statistical computing languages: MATLAB and Python (example code in Figure \ref{fig:code}). The code for both implementations is available at \url{https://github.com/nrdg/fracridge} and released under an OSI-approved, permissive open-source license to facilitate its broad use. In both MATLAB and Python, we used broadcasting to rapidly perform computations over multiple dimensions of arrays.

There are two potential performance bottlenecks in the code. One is the SVD step which is expensive both in terms of memory and computation time. This step is optimized by observing that for cases in which $d>p$ (the number of data points is larger than the number of parameters), we can replace the singular values of $X$ by the square roots of the singular values of $X^tX$, which is $p$ by $p$ and therefore requires less memory for the SVD computation. The other potential performance bottleneck is the interpolation performed for each target. To optimize this step, we used fast interpolation functions that assume sorted inputs.

\subsubsection{Matlab}

The MATLAB implementation of FRR relies only on core MATLAB functions and a fast implementation of linear interpolation \citep{interp1qr}, which is copied into the fracridge source code, together with its license, which is compatible with the fracridge license. The MATLAB implementation includes an option to automatically standardize predictors (either center or also scale the predictors) before regularization, if desired.

\subsubsection{Python}
The Python implementation of FRR depends on Scipy \citep{Virtanen2020-mf} and Numpy \citep{Van_der_Walt2011-em}. 
The object-oriented interface provided conforms with the API of the popular Scikit-Learn library \citep{pedregosa2011scikit, buitinck2013api}, including automated tests that verify compliance with this API. Unit tests are implemented using pytest \citep{pytest5.4.1}. Documentation is automatically compiled using sphinx, with sphinx-gallery examples \citep{sphinx_gallery_2020}. The Python implementation also optionally uses Numba \citep{Lam2015-lr} for just-in-time compilation of a few of the underlying numerical routines used in the implementation. This functionality relies on an implementation provided in the hyperlearn library \citep{hyperlearn} and copied into the fracridge source-code, together with its license, which is compatible with the fracridge license. In addition to its release on GitHub, the software is available to install through the Python Package Index (PyPI) through the standard Python Package Installer (\texttt{pip install fracridge}). For Python, we did not implement standardization procedures, as those are implemented as a part of Scikit-Learn.

\begin{figure}[ht!]
\noindent\begin{minipage}{.5\textwidth}

\begin{lstlisting}[language=Matlab, title=Matlab]{Name}

y = randn(100,1);
X = randn(100,10);

% Calculate coefficients with naive OLS
coef = inv(X'*X)*X'*y;

% Call the fracridge function:
[coef2, alpha] = fracridge(X, 0.3, y);

% Calculate coefficients with naive RR
alphaI = alpha*eye(size(X, 2));
coef3 = inv(X'*X + alphaI)*X'*y;

norm(coef)
norm(coef2)
norm(coef2) ./ norm(coef)
norm(coef2-coef3)


\end{lstlisting}
\end{minipage}\hfill
\begin{minipage}{.5\textwidth}
\begin{lstlisting}[language=Python, title=Python]{Name}

import numpy as np
from numpy.linalg import inv, norm
from fracridge import fracridge
from fracridge import FracRidge

y = np.random.randn(100)
X = np.random.randn(100, 10)

# Calculate coefficients with naive OLS
coef = inv(X.T @ X) @ X.T @ y

# Call fracridge function:
coef2, alpha = fracridge(X, y, 0.3)

# Calculate coefficients with naive RR
alphaI = alpha * np.eye(X.shape[1])
coef3 = inv(X.T @ X + alphaI) @ X.T @ y

print(norm(coef))
print(norm(coef2))
print(norm(coef2) / norm(coef))
print(norm(coef2 - coef3))

# sklearn-compatible object-oriented API:
fr = FracRidge(fracs=0.3)
fr.fit(X, y)
coef_oo = fr.coef_
alpha_oo = fr.alpha_
print(norm(coef_oo) / norm(coef))

\end{lstlisting}
\end{minipage}
\caption{Code examples. Left: MATLAB examples that demonstrate the software API and correctness of the implementation. Right: Python examples demonstrate a similar API and correctness. Python examples include the Sckit-Learn-compatible API}
    \label{fig:code}
\end{figure}

\subsection{Simulations}
Numerical simulations were used to characterize FRR and compare it to a heuristic approach for hyperparameter selection. Simulations were conducted using the MATLAB implementation. We simulated two simple regression scenarios. The number of data points ($d$) was 100, and the number of predictors ($p$) was either 5 or 100. In each simulation, we first created a design matrix X ($d, p$) using the following procedure: (i) generate normally distributed values for X, (ii) induce correlation between predictors by selecting two predictors at random, setting one of the predictors to the sum of the two predictors plus normally distributed noise, and repeating this procedure $2p$ times, and (iii) z-scoring each predictor. Next, we created a set of ``ground truth'' coefficients $\beta$ with dimensions ($p, 1$) by drawing values from the normal distribution. Finally, we simulated responses from the model ($y=X\beta$) and added normally distributed noise, producing a target variable $y$ with dimensions ($d, 1$).

Given design matrix $X$ and target $y$, cross-validated regression was carried out. This was done by splitting $X$ and $y$ into two halves (50/50 training/testing split), solving ridge regression on one half (training) and evaluating generalization performance of the estimated regression $\beta$ weights on the other half (testing). Performance was quantified using the coefficient of determination ($R^2$). For standard ridge regression, we evaluated a grid of $\alpha$ values that included 0 and ranged from $10^{-4}$ through $10^{5.5}$ in increments of $0.5~log_{10}$ units. For FRR, we evaluated a range of fractions $\gamma$ from 0 to 1 in increments of 0.05. Thus, the number of hyperparameter values was $f=21$ in both cases.

The code that implements these simulations is available in the \href{https://github.com/nrdg/fracridge/blob/0.4/examples/paper_figures/Fig1.m}{``examples'' folder of the software}.

\subsection{Performance benchmark}

To characterize the performance of fractional ridge regression (FRR) and standard ridge regression (SRR) approaches, a set of numerical benchmarks was conducted using the MATLAB implementation. A range of regression scenarios were constructed. In each experiment, we first constructed a design matrix X ($d, p$) consisting of values drawn from a normal distribution. We then created ``ground truth'' coefficients $\beta$ ($p, t$) also by drawing values from the normal distribution. Finally, we generated a set of data $Y$ ($d, t$) by predicting the model response ($y=X\beta$) and adding zero-mean Gaussian noise with standard deviation equal to the standard deviation of the data from each target variable. Different levels of regularization ($f$) were obtained for SRR by linearly spacing $\alpha$ values on a $log_{10}$ scale from $10^{-4}$ to $10^5$ and for FRR by linearly spacing fractions from 0.05 to 1 in increments of 0.05.

Two versions of SRR were implemented and evaluated. The first version (na\"{i}ve) involves a separate matrix (pseudo-)inversion for each hyperparameter setting desired. The second version (rotation-based) involves using the SVD decomposition method described above (see section \ref{sec:background}, specifically equation \ref{eq:rr_coeff}).

All simulations were run on an Intel Xeon E5-2683 2.10 Ghz (32-core) workstation with 128 GB of RAM, a 64-bit Linux operating system, and MATLAB 8.3 (R2014a). Execution time was logged for model fitting procedures only and did not include generation of the design matrix or the data. Likewise, memory requirements were recorded in terms of additional memory usage during the course of model fitting (i.e. zero memory usage corresponds to the total memory usage just prior to the start of model fitting). Benchmarking results were averaged across 15 independent simulations to reduce incidental variability.

The code that implements these benchmarks is available in the \href{https://github.com/nrdg/fracridge/blob/0.4/examples/paper_figures/Fig2.m}{``examples'' folder of the software}.

\subsection{Brain Magnetic Resonance Imaging data}
Brain functional Magnetic Resonance Imagine (fMRI) data were collected in a 7 Tesla MRI instrument, at a spatial resolution of $1.8~mm$ and a temporal resolution of $1.6~s$ and using a matrix size of [81 104 83]. This yielded a total of 783,432 voxels. Over the course of 40 separate scan sessions, a participant viewed 10,000 distinct images (3 presentations per image) while fixating a small dot placed at the center of the images (see Figure 3A). The images were $8.4\deg$ by $8.4\deg$ in size. Each image was presented for $3~s$ and was followed by a $1~s$ gap. Standard pre-processing steps were applied to the fMRI data to remove artifacts due to head motion and other confounding factors. To deal with session-wise nonstationarities, response amplitudes of each voxel were z-scored within each scan session. Responses were then concatenated across sessions and averaged across trials of the same image, and then a final z-scoring of each voxel's responses was performed.

A regression model was used to predict the response observed from a voxel in terms of local contrast present in the stimulus image. In the model, the stimulus image is pre-processed by taking the original color image (425 pixels by 425 pixels by 3 RGB channels), converting the image to grayscale, gridding the image into 25 by 25 regions, and then computing the standard deviation of luminance values within each grid region (Figure \ref{fig:realdata}B). This produced 625 predictors, each of which was then z-scored. The design matrix $X$ has dimensionality 10,000 images by 625 stimulus regions, while $Y$ has dimensionality 10,000 images by 783,432 voxels.

Cross-validation was carried out using a 80/20 training/testing split. For standard ridge regression, we evaluated a grid of alpha values that included 0 and ranged from $10^{-4}$ to $10^{5.5}$ in increments of $0.5~log_{10}$ units. For fractional ridge regression, we evaluated a range of fractions from 0 to 1 in increments of 0.05. Cross-validation performance was quantified in terms of variance explained on the test set using the coefficient of determination ($R^2$).

The code that implements these benchmarks is available in the \href{https://github.com/nrdg/fracridge/blob/0.4/examples/paper_figures/Fig3.m}{``examples'' folder of the software}.

\section{Results}

\subsection{Fractional ridge regression achieves the desired outcomes}
\label{sec:correctness}

\begin{figure}[ht!]
    \includegraphics[width=0.98\textwidth]{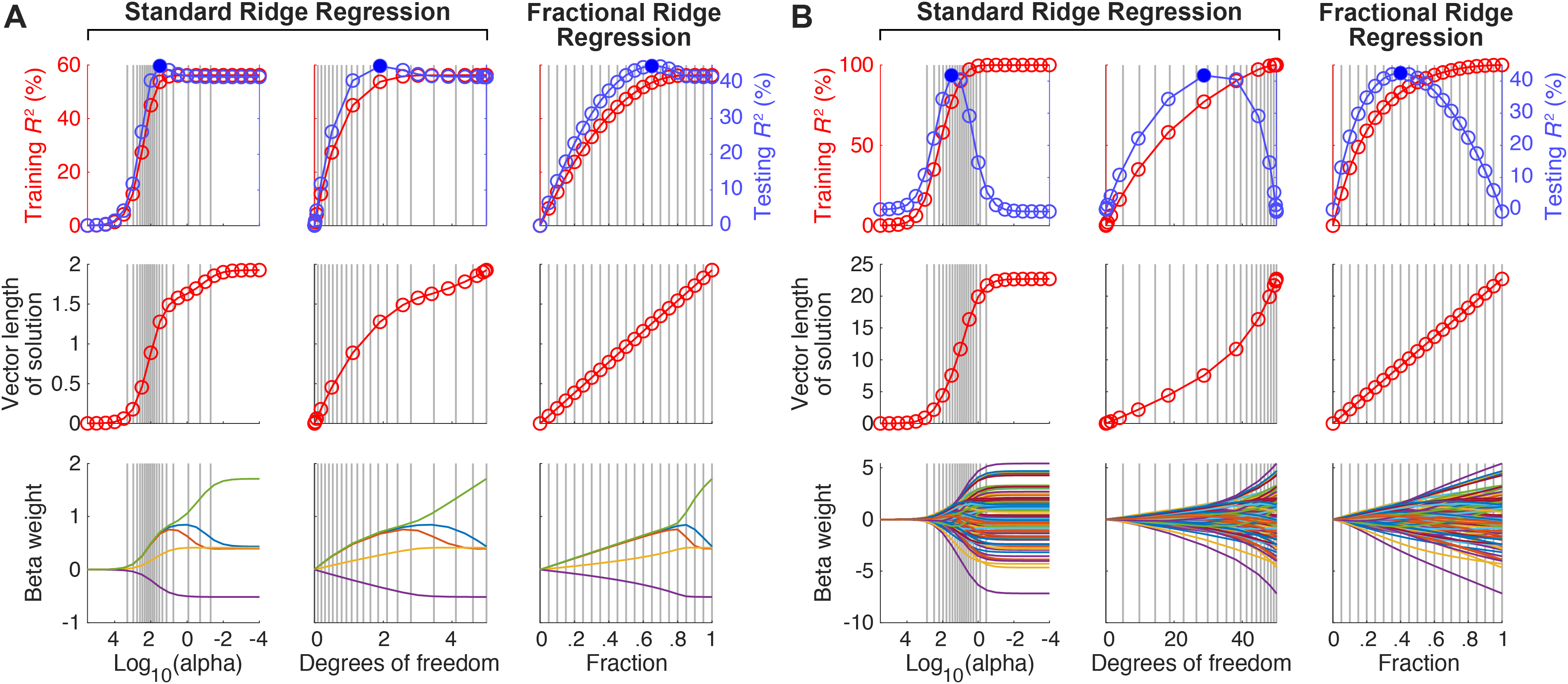}
    \caption{{\bf Fractional ridge regression (FRR) achieves desired outcomes}. (A) Example regression scenario ($d = 100$, $p = 5$). The first two columns show the results of standard ridge regression in which log-spaced $\alpha$ values are used to obtain different levels of regularization. Whereas the first column shows results as a function of $log_{10}(\alpha)$, the second column shows results as a function of $\alpha$ values converted to effective degrees of freedom (see Methods). The third column shows the results of fractional ridge regression in which different regularization levels are achieved by requesting specific fractional L2-norm ($\gamma$). Solid blue dots mark peak cross-validation performance. Vertical gray lines in the third column indicate regression solutions obtained by the FRR method (requested fractions range from 0 to 1 in increments of 0.05). The corresponding locations of these regression solutions in the first and second columns are also shown using vertical gray lines. The bottom row shows coefficient paths, i.e., the values of $\beta$ as a function of $log_{10}(\alpha)$, degrees of freedom, or fraction $\gamma$. (B) Example regression scenario ($d = 100$, $p = 100$). Same format as panel A. Notice that in both scenarios, only the FRR method achieves regression solutions whose L2-norms increase linearly, with gradually changing coefficient paths.}
    \label{fig:correctness}
 \end{figure}

In simulations, we demonstrate that the fractional ridge regression (FRR) algorithm accurately produces the desired fractions $\gamma$ (Figure \ref{fig:correctness} A,B second row, right column in each). We compare the results of FRR to results of standard ridge regression (SRR), in which a commonly-used heuristic is used to select $\alpha$ values (log-spaced values spanning a large range). For the SRR approach, we find that the fractional L2-norm is very small and virtually indistinguishable for large values of $\alpha$, and is very similar to the OLS solution (fractional L2-norm approximately 1) for several small values of $\alpha$ (Figure \ref{fig:correctness} A, B second row, left column). In addition, cross-validation accuracy is indistinguishable for many of the values of $\alpha$ evaluated in SRR. Only very few values of $\alpha$ produce cross-validated $R^2$ values that are similar to the value provided by the best $\alpha$ (Figure \ref{fig:correctness} A, B first row, left column).

The SRR results can also be re-represented using effective degrees of freedom (DOF; Figure \ref{fig:correctness} A, B first row, middle column): several values of $\alpha$ result in essentially the same number of DOF, because these values are either much larger than the largest singular value or much smaller than the smallest singular value of $X$. In contrast to SRR, FRR produces a nicely behaved range of cross-validated $R^2$ values and dense sampling around the peak $R^2$.

Another line of evidence highlighting the diversity of the solutions provided by FRR is given by inspecting coefficient paths: in the log-spaced case, coefficients start very close to 0 (for high $\alpha$) and rapidly increase (for lower $\alpha$). Even when re-represented using DOF, the coefficient paths exhibit some redundancy. In contrast, FRR provides more gradual change in the coefficient paths, indicating that this approach explores the space of possible coefficient configurations more uniformly. Taken together, these analyses demonstrate that FRR provides a more useful range of regularization levels than SRR.

 \subsection{FRR is computationally efficient}
\label{sec:results_benchmark}
A question of relevance to potential users of FRR is whether using the method incurs significant computational cost. We compare FRR to two alternative approaches. The first approach is a na\"{i}ve implementation of the matrix inversion specified in equation \ref{eq:rr}, in which the Moore-Penrose pseudo-inverse (implemented as \texttt{pinv} in Matlab and \texttt{numpy.linalg.pinv} in Python) is performed independently for each setting of hyperparameter $\alpha$. The second approach takes advantage of the computational expedience of the SVD-based approach: instead of a matrix inversion for each $\alpha$ value, a single SVD is performed, a transformation (rotation) is applied to the data, and different values of $\alpha$ are plugged into equation \ref{eq:rr_coeff} to compute the regression coefficients. This approach comprises a subset of the operations taken in FRR. Therefore, it represents a lower bound in terms of computational requirements.

\begin{figure}[ht!]
    \includegraphics[width=0.98\textwidth]{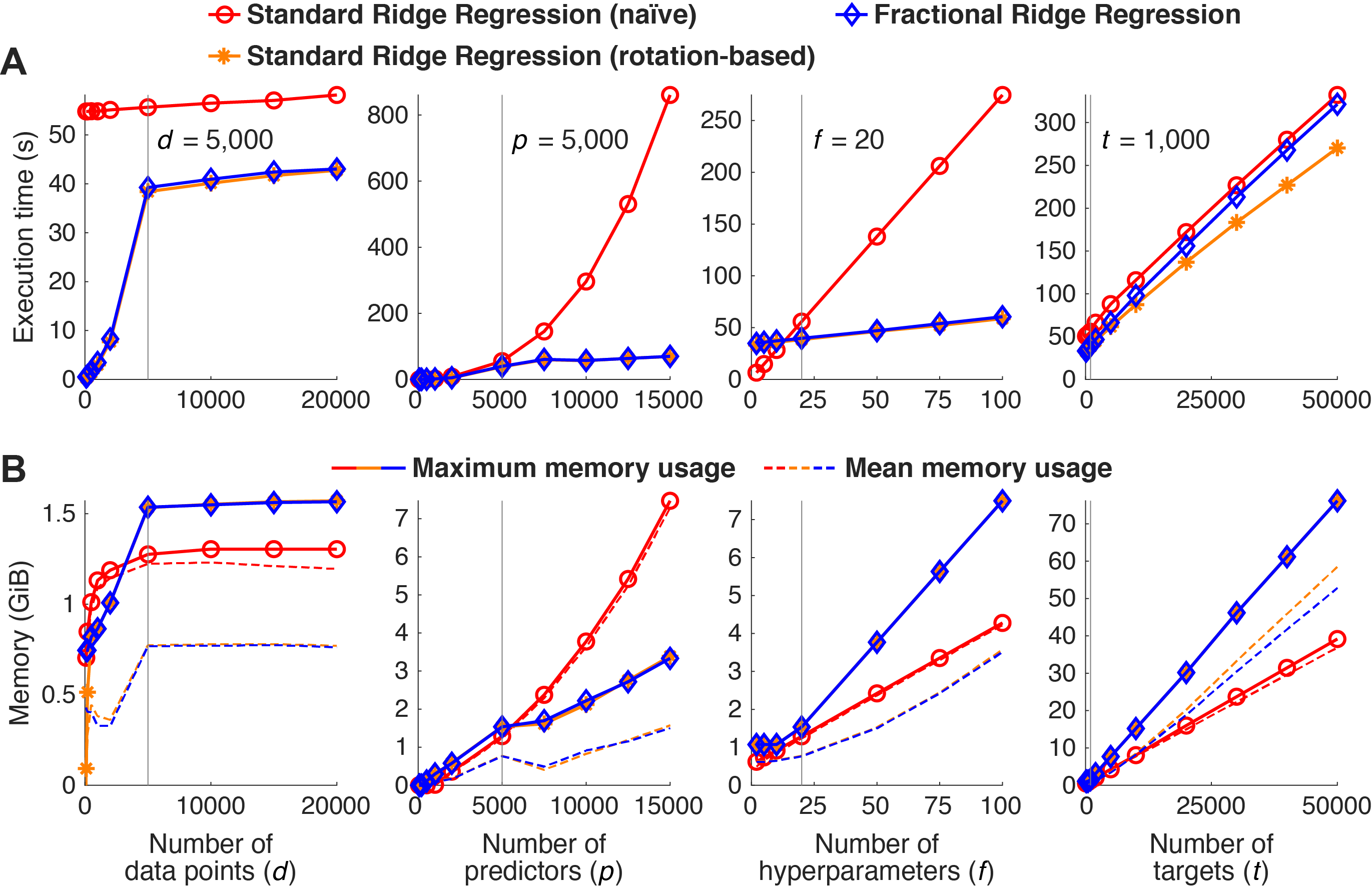}
    \caption{{\bf Computational efficiency}. We benchmarked different methods for performing ridge regression: (1) a na\"{i}ve implementation of standard ridge regression (involving log-spaced $\alpha$ values) in which matrix inversion is performed for each $\alpha$ value, (2) an implementation of standard ridge regression in which solutions are computed in a rotated space based on singular value decomposition of the design matrix, and (3) the FRR method. Starting from a base case ($d = 5,000$, $p = 5,000$, $f = 20$, $b = 1,000$; parameter settings marked by vertical lines), we systematically manipulated $d$, $p$, $f$, and $b$ (columns one through four, respectively). (A) Execution time. The execution time of each method is shown in seconds. (B) Memory usage. The maximum memory usage of each method is shown as a solid line, whereas the time-averaged memory usage is shown as a dotted line. Overall, FRR is quite fast and has relatively modest memory requirements.}
    \label{fig:benchmark}
 \end{figure}

Through systematic exploration of different problem sizes, we find that FRR performs quite favorably. FRR differs from the rotation-based approach only slightly with respect to execution-time scaling in the number of data points (Figure \ref{fig:benchmark}A, left column), in the number of parameters (Figure \ref{fig:benchmark}A, right column), and in $f$, the number of hyperparameter values considered (Figure \ref{fig:benchmark}A, third column ). The na\"{i}ve matrix-inversion approach is faster than both SVD-based approaches (FRR and rotation-based) for $f<20$, but rapidly becomes much more costly for values above 20. This approach also scales rather poorly for $p>5,000$.

In terms of memory consumption, the mean and maximum memory usage are very similar for FRR, the na\"{i}ve and rotation-based solutions. These results suggest that for each of these approaches, the matrix inversion (for the na\"{i}ve implementation of SRR) or the SVD (for FRR and the SVD-based SRR) represents the main computational bottleneck. Importantly, despite the fact that FRR uses additional gridding and interpolation steps, it does not perform substantially worse than either of the other approaches.

\subsection{Application of FRR on real-world data}
\label{sec:fmri}
\begin{figure}[ht!]
    \includegraphics[width=0.65\textwidth]{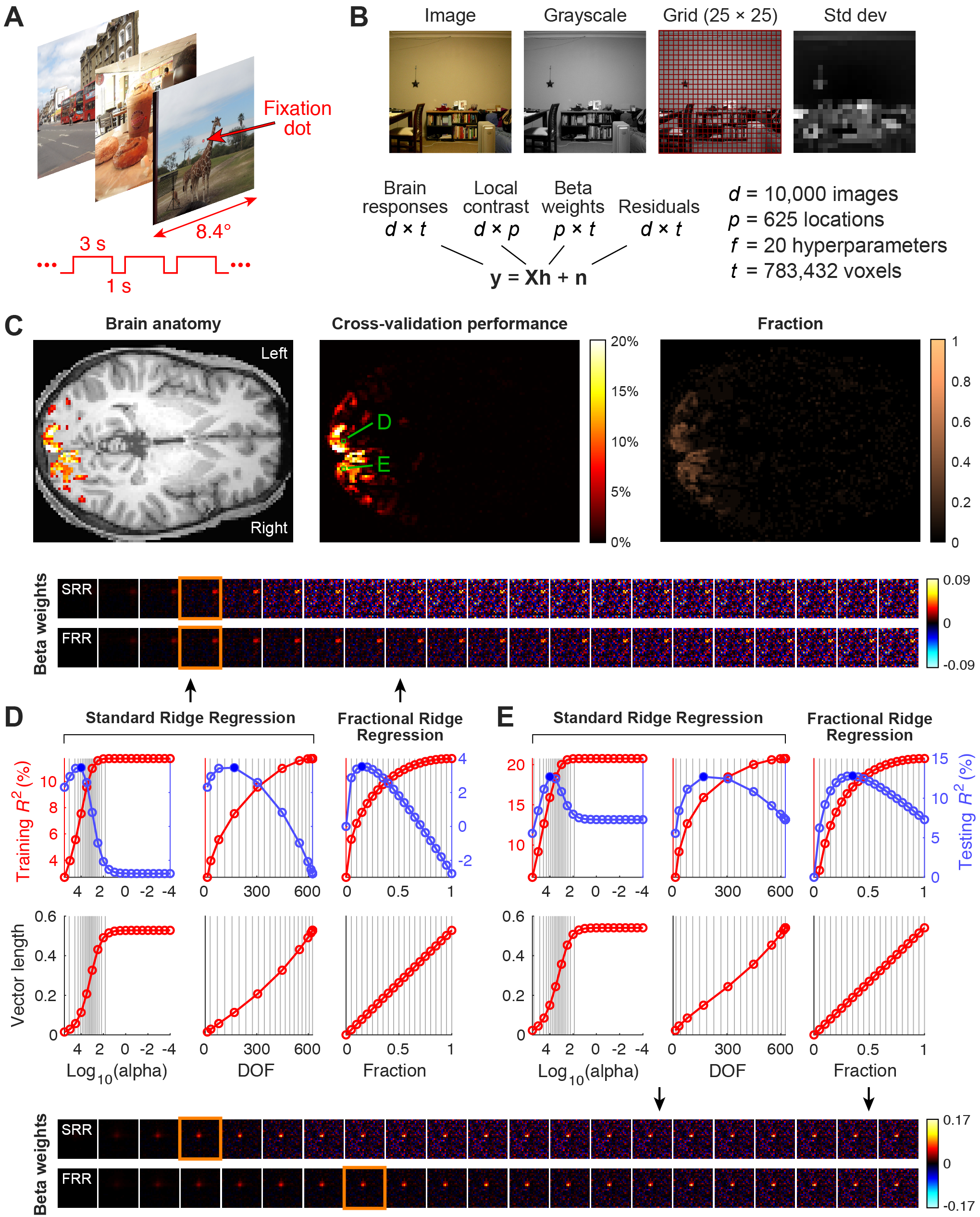}
    \caption{{\bf Demonstration on real-world data}. (A) Visual fMRI experiment. Functional MRI measurements of brain activity were collected from a human participant while s/he viewed a series of natural images (10,000 distinct images presented three times each). (B) Model of brain activity. Images were converted to grayscale and gridded, and then standard deviation of luminance values within each grid element was calculated. This produced measures of local contrast. Brain responses at every voxel were modeled using a weighted sum of local contrast. (C) Results obtained using FRR. Cross-validated performance (variance explained) achieved by the model is shown for an axial brain slice (middle). These results are thresholded at 5\% and superimposed on an image of brain anatomy for reference (left). The fraction ($\gamma$) corresponding to the best cross-validation performance is also shown (right). (D) Detailed results for one voxel (see green squares in panel C). The main plots that depict training and testing performance and L2-norm are in the same format as Figure 1. The inset illustrates coefficient solutions for different regularization levels. The orange box highlights the regularization level producing highest cross-validation performance. (E) Detailed results for a second voxel. Same format as panel D.}
    \label{fig:realdata}
\end{figure}

To demonstrate the practical utility of FRR, we explore its application in a specific scientific use-case. Data from a functional Magnetic Resonance Imaging (fMRI) experiment were analyzed with FRR and the results of this analysis were compared to a standard ridge regression (SRR) approach where $\alpha$ values are selected using a log-spaced heuristic. Different parts of the brain process different types of information, and a large swath of the cerebral cortex is known to respond to visual stimulation. Experiments that combine fMRI with computational analysis provide detailed information about the responses of different parts of the brain \citep{wandell2015computational}. In the experiments analyzed here, a series of images are shown and the MRI blood-oxygen-level-dependent (BOLD) signal is recorded in a sampling grid of voxels throughout the brain (Figure \ref{fig:realdata}A). In the cerebral cortex, each voxel contains hundreds of thousands of neurons. If these neurons respond vigorously to the visual stimulus presented, the metabolic demand for oxygen in that part of cortex will drive a transient increase in oxygenated blood in that region, and the BOLD response will increase. Thus, a model of the BOLD response tells us about the selective responses of neurons in each voxel in cortex.

Because neurons in parts of the cerebral cortex that respond to visual stimuli are known to be particularly sensitive to local contrast, we model responses with respect to the standard deviation of luminance in each region of the image, rather than the luminance values themselves (Figure \ref{fig:realdata}B). In the model, $Y$ contains brain responses where each target (column) represents the responses in a single voxel. Each row contains the response of all voxels to a particular image. The design matrix $X$ contains the local contrast in every region of the image, for every image. This means that the coefficients $\beta$ represent weights on the stimulus image and indicate each voxel's spatial selectivity -- i.e., the part of the image to which the voxel responds \citep{Wandell2015-jf}. Therefore, one way to visualize $\hat{\beta}$ is to organize it according to the two-dimensional layout of the image (Figure \ref{fig:realdata}C\&D, bottom two rows).

Using FRR, we fit the model to voxel responses, and find robust model performance in the posterior part of the brain where visual cortex resides (left part of the horizontal slice presented in the top row of Figure \ref{fig:realdata}C). The performance of the model can be observed in either the cross-validated $R^2$ values (Figure \ref{fig:realdata}C, top row, left and middle panels) or the value of $\gamma$ corresponding to the best cross-validated $R^2$. For example, we can focus on the two voxels highlighted in the middle panel of the top row in Figure \ref{fig:realdata}C. One voxel, whose characteristics are further broken down in Figure \ref{fig:realdata}D has lower cross-validated $R^2 = 4\%$ and requires stronger regularization ($\gamma = 0.15$). The spatial selectivity of this voxel's responses becomes very noisy at large $\gamma$ values and $R^2$ approaches 0. On the other hand, the voxel in Figure \ref{fig:realdata}E has a higher $\gamma = 0.35$ and a higher cross-validated $R^2=13\%$. Moreover, this voxel appears more robust with higher values of $\gamma$ producing less spatially noisy results. The map of $R^2$ and $\gamma$ illustrated in Figure \ref{fig:realdata}C show that these trends hold more generally: voxels with more accurate models require less regularization.

\section{Discussion}

The main theoretical contribution of this work is a novel approach to hyperparameter specification in ridge regression. Instead of the standard approach in which a heuristic range of values for hyperparameter $\alpha$ are evaluated for their accuracy, the fractional ridge regression (FRR) approach focuses on achieving specific fractions for the L2-norms of the solutions relative to the L2-norm of the unregularized solution. In a sense, this is exactly in line with the original spirit of ridge regression, which places a penalty on the L2-norm of the solution. The main practical contribution of this work is the design and implementation of an efficient algorithm to solve FRR and validation of this algorithm on simulated and empirical data. Overall, we suggest that FRR can serve as a default approach to solving ridge regression.

\subsection{The benefits of FRR}

\begin{enumerate}
    \item \textbf{Theoretically-motivated and principled}. The results in section \ref{sec:correctness} demonstrate that the theoretical motivation described in the Methods holds in practice. Our implementation of FRR produces ridge regression solutions that have predictable and tuneable fractional L2-norm. 
    \item \textbf{Statistically efficient}. Each fraction level returned by FRR produces $\beta$ values that are distinctly different. This avoids the common pitfall in the log-spaced approach whereby computation is wasted on several values of $\alpha$ that all over-regularize or under-regularize. When used with a range of $\gamma$ values from 0 to 1, the solution that minimizes cross-validation error is guaranteed to exist within this range (although it may lie in between two of the obtained solutions).
    \item \textbf{Computationally efficient}. We show that our implementation of FRR requires memory and computational time that are comparable to a na\"{i}ve ridge regression approach and to an approach that uses SVD but relies on preset $\alpha$ values. SVD-based approaches (including FRR) scale linearly in $f$, with compute-time scaling better than na\"{i}ve RR in the $f>20$ regime. In practice, we have found that $f=20$ evenly distributed values between 0 and 1 provide sufficient coverage for many problems. But the linear scaling implies that sampling more finely would not be limiting in cases where additional precision is needed.
    
    \item \textbf{Interpretable}. FRR uses $\gamma$ values that represent scaling relative to the L2-norm of the OLS solution. This allows FRR results to be compared across different targets within a dataset. This is exemplified in section \ref{sec:fmri}, in which results from an fMRI experiment are interpreted both in light of cross-validated $R^2$ and the optimal $\gamma$ that leads to the best cross-validated $R^2$. Moreover, regularization in different datasets and for different models (e.g., different settings of $X$) can be compared to each other as being stronger or weaker. The optimal regularization level can be informative regarding the signal-to-noise of a particular target or about the level of collinearity of the design matrix (which both influence the optimal level of regularization). FRR increases the interpretability of ridge regression, because instead of an unscaled, relatively inscrutable value of $\alpha$, we receive a scaled, relatively interpretable value. Based on a recently proposed framework for interpretability in machine learning methods \citep{Murdoch2019-ax}, we believe that this kind of advance improves the descriptive accuracy of ridge regression. 
    
    \item \textbf{Automatic}. Machine learning algorithms focus on automated inferences, but many machine learning algorithms still require substantial manual tuning. For example, if the range of $\alpha$ values used is not sufficient, users of ridge regression may be forced to explore other values. This is impractical in cases in which thousands of targets are analyzed and multiple models evaluated. Thus, FRR contributes to the growing field of methods that aim to automate machine learning methods \citep{Zoller2019-il, Tuggener2019-in}. These methods all aim to remove the burden of manual inspection and tuning of machine learning. A major benefit of FRR is therefore practical in nature: Because FRR spans the dynamic range of effects that ridge regression can provide, using FRR guarantees that the time taken to explore hyperparameter values is well spent. Moreover, the user does not have to spend time speculating what $\alpha$ values might be appropriate for a given problem (e.g. is $10^{4}$ sufficiently high?).
    
    \item \textbf{Implemented in usable open-source software}. We provide code that is well-documented, thoroughly tested, and easy to use: \url{https://github.com/nrdg/fracridge}. The software is available in two popular statistical programming languages: MATLAB and Python. The Python implementation provides an object-oriented interface that complies with the popular Scikit-Learn library \citep{pedregosa2011scikit, buitinck2013api}.

\end{enumerate}

\subsection{Limitations}

One limitation of FRR is that a heuristic approach is used within the algorithm to generate the grid of $\alpha$ values used for interpolation (see section \ref{sec:methods_algorithm} for details). Nonetheless, the interpolation results are quite accurate, and costly computations are carried out only for final desired $\alpha$ values. Another limitation is that the $\alpha$ value that corresponds to a specific $\gamma$ may be different for different targets and models. If there are theoretical reasons to retain the same $\alpha$ across targets and models, the FRR approach is not appropriate. But this would rarely be the case, as $\alpha$ values are usually not directly interpretable. Alternatively, FRR can be used to estimate values of $\alpha$ on one sample of the data (or for one model) and these values of $\alpha$ can then be used in all of the data (or all models).

Finally, the FRR approach is limited to ridge regression and does not generalize easily to other regularization approaches. The Lasso \citep{tibshirani1996lasso} provides regression solutions that balance least-squares minimization with the L1-norm of the coefficients, rather than the L2-norm of the coefficients. The Lasso approach has several benefits, including results that are more sparse and potentially easier to interpret. Similarly, Elastic Net \citep{zou2005regularization} uses both L1- and L2-regularization, potentially offering more accurate solutions. But because the computational implementation of these approaches differs quite substantially from ridge regression, the approach presented in this paper does not translate easily to these methods. Moreover, while these methods allow regularization with a non-negativity constraint on the coefficients, this constraint is not easily incorporated into L2-regularization. On the other hand, a major challenge that arises in L1-regularization is computational time: most algorithms operate for one target at a time and incur substantial computational costs, and scaling such algorithms to the thousands of targets in large-scale datasets may be difficult.

\subsection{Future extensions}

An important extension of the present work would be an implementation of these ideas in additional statistical programming languages, such as the R programming language, which is very popular for use in statistical analysis of data from many different domains. One of the most important tools for regularized regression is the glmnet software package which was originally implemented in the R programming language \citep{friedman2009glmnet} and has implementations in MATLAB \citep{qian2013glmnet} and Python \citep{balakumar2016glmnet}. The software also provides tools for analysis and visualization of coefficient paths and of the effects of regularization on cross-validated error. The R glmnet vignette \citep{hastie2014glmnet} demonstrates the use of these tools. In addition to identifying the $\alpha$ value that minimizes cross-validation error, glmnet also identifies the $\alpha$ which gives the most regularized model such that the cross-validated error is within one standard error of the minimum cross-validated error. This approach acknowledges that there is some error in selecting $\alpha$ and chooses to err on the side of a more parsimonious model \citep{friedman2010regularization}. Future extensions of FRR could implement this heuristic.

\acks{The authors would like to thank Noah Simon for helpful discussions. AR was funded through a grant from the Gordon \& Betty Moore Foundation and the Alfred P. Sloan Foundation to the University of Washington eScience Institute, through NIH grants 1RF1MH121868-01 (PI: AR) from the National Institute for Mental Health and 5R01EB027585-02 (PI: Eleftherios Garyfallidis, Indiana University) from the National Institute for Biomedical Imaging and Bioengineering and through NSF grants 1934292 (PI: Magda Balazinska, University of Washington). KK was supported by NIH P41 EB015894. Collection of MRI data was supported by NIH S10 RR026783 and the W.M. Keck Foundation.}


\newpage

\vskip 0.2in
\bibliography{paper}

\end{document}